\documentclass[conference]{IEEEtran}
\IEEEoverridecommandlockouts

\usepackage{cite}
\usepackage{amsmath,amssymb,amsfonts}
\usepackage{algorithmic}
\usepackage{graphicx}
\usepackage{textcomp}
\usepackage{xcolor}
\usepackage[linesnumbered,ruled,lined]{algorithm2e}
\usepackage{balance}
\usepackage{multirow}
\usepackage{dblfloatfix}
\usepackage{bbding}
\usepackage{lipsum}
\usepackage{hyperref}
\usepackage{makecell}
\usepackage{array}
\usepackage{booktabs}
\usepackage{color}
\usepackage{setspace}
\usepackage{diagbox}
\usepackage{pifont}
\usepackage[
 font=scriptsize
 ]{subfig}

\def\BibTeX{{\rm B\kern-.05em{\sc i\kern-.025em b}\kern-.08em
    T\kern-.1667em\lower.7ex\hbox{E}\kern-.125emX}}
\begin{document}

\title{What Features Influence Impact Feel? \\A Study of Impact Feedback in Action Games}

\author{\IEEEauthorblockN{Zhonghao Lin}
\IEEEauthorblockA{\textit{School of Science and Engineering} \\
\textit{The Chinese Univerisity of Hong Kong, Shenzhen}\\
Shenzhen, China \\
zhonghaolin@link.cuhk.edu.cn}
\and
\IEEEauthorblockN{Haihan Duan}
\IEEEauthorblockA{\textit{School of Science and Engineering} \\
\textit{The Chinese Univerisity of Hong Kong, Shenzhen}\\
Shenzhen, China \\
haihanduan@link.cuhk.edu.cn}
\and
\IEEEauthorblockN{Zikai Alex Wen}
\IEEEauthorblockA{\textit{Computational Media and Arts Thrust} \\
\textit{The Hong Kong University of Science and Technology (Guangzhou)}\\
Guangzhou, China \\
zikaiwen@ust.hk}
\and
\IEEEauthorblockN{Wei Cai*\thanks{*Wei Cai is the corresponding author (caiwei@cuhk.edu.cn).}}
\IEEEauthorblockA{\textit{School of Science and Engineering} \\
\textit{The Chinese Univerisity of Hong Kong, Shenzhen}\\
Shenzhen, China \\
caiwei@cuhk.edu.cn}
}

\maketitle

\begin{abstract}
Making the hit effect satisfy players is a long-standing problem faced by action game designers. However, no research has systematically analyzed which game design elements affect such game feel. There is not even a term to describe it. So, we proposed to use \textit{impact feel} to describe the player's feeling when receiving juicy impact feedback. After collecting player's comments on action games from \textit{Steam}'s top seller list, we trained a natural language processing (NLP) model to rank action games with their performance on impact feel. We presented a 19-feature framework of impact feedback design and examined it in the top eight and last eight games. We listed an inventory of the usage of features and found that hit stop, sound coherence, and camera control may strongly influence players' impact feel. A lack of dedicated design on one of these three features may ruin players' impact feel. Our findings might become an evaluation metric for future studies.
\end{abstract}


\section{Introduction}
In March 2022, the newly released soulslike action game \textit{Elden Ring}\footnote{https://en.bandainamcoent.eu/elden-ring/elden-ring} reached a new peak 950K concurrent players on \textit{Steam}\footnote{https://store.steampowered.com/}, which ranked the 6th highest among all of the platform's games. The phenomenon revealed that the growing enthusiasm players were devoted to action games. Impact feedback design is a critical factor that contributes to the success of video games. For action games that focus the gameplay on providing a tasty fighting experience, good impact feedback is indispensable. There are many successful action games winning players' appreciation for providing awesome feeling when receiving impact feedback, such as  
\textit{Street Fighter}\footnote{https://streetfighter.com/en/}, \textit{Monster Hunter}\footnote{https://www.monsterhunter.com/}, and \textit{God of War}\footnote{https://www.playstation.com/zh-hans-hk/games/god-of-war/} Series, 
while there are also games that come under fire for the awful feeling of impact.

However, to the best of our knowledge, no word explicitly defines this subtle feeling. Therefore, we proposed the word impact feel, a sub-genre of game feel, to describe the satisfying feeling when receiving the impact of hits. It is the visceral emotion derived from a well-designed impact feedback system that requires juiciness in visual, auditory, and haptic feedback.
 
How to improve impact feel is a significant problem for action game designers. There exist plenty of technologies applied in impact feedback system to provide juiciness. Hundreds of website articles have described how these technologies are implemented in games. However, the significant part ``selection of games" is mainly based on empirical judgment, which is not convincing for research purposes. 
Unlike passionate players and game developers online, only a few researchers target impact feel.
Martin \cite{pichlmair} presented several technologies that contribute to making an attack look more potent in his survey of game feel, such as hit stop and slow motion. However, his evidence came from the empirical analysis from a game designer's blog \cite{song}. Therefore, we aimed to carry out a data-driven analysis to identify features influencing the impact feel.

To achieve the target, we first selected popular fighting games and ranked them for the performance of impact feel by NLP. By exploring the difference between games with high and low ranks, we discovered features that influence players' feeling of impact. Our work contributes to these aspects: First, we proposed the word impact feel to describe players' game feel when receiving impact feedback. It fulfills the requirement for a concise and uniform definition on this area and makes it easy to study. Second, we collected game comments from \textit{Steam}, manually annotate, and constructed a data set of 5000 comments about game's impact feel. On this basis, we provided a data-driven ranking for popular action games with their performance on impact feel by an NLP model. Based on previous work on juicy design \cite{hicks1}, we presented a 19-feature framework in impact feedback design that applies to popular action games. By experiment, we identified that for action games, as long as there is a lack of meticulous design in hit stop, sound coherence, and camera control, the performance of impact feel will be unsatisfactory.

\section{Related Work}
\textbf{Game Feel and Impact Feel.} The word game feel is firstly proposed by Swink \cite{Swink} in his book \textit{Game Feel}, described as ``real time control of virtual objects in a simulated space, with interactions emphasized by polish". Several researchers have discussed their opinions about game feel in the following years. Martin \cite{pichlmair} proposes the latest perspective; he argues that game feel should ``focus on the player and the design of their interaction with the game." As the narrow set of game feel focusing on the fighting process, the feeling of impact is a critical element for action games. Methodologies 
used in successful action games are discussed frequently on some websites and online forums, such as \textit{Game Developer} and \textit{Reddit}. There are several expressions describing the feeling of impact in previous articles, such as ``combat impact" \cite{song}, ``the feeling of hit" \cite{moon}, and ``hit effect" \cite{peyman}. Most expressions lack academic reference, and the diversity makes this subtle feeling difficult to search and study. Therefore, the requirement for a unified and convincing definition is urgent. In 2021, the word ``impact feedback visualization" is proposed in a survey of designing game feel \cite{pichlmair}. We derive the word ``impact feel" to describe the feeling when receiving ``impact feedback".
Several technologies have been discussed for the influence on impact feel such as hit stop, slow motion and special effects. 
Daniels \cite{daniels}, the former combat designer of \textit{God of War},  presents the importance of hit stop in making a game feel better.
Song \cite{song} provides an overview of how to model the combat impact in action games, including animation, camera and special effects. 
Massoud \cite{peyman} discusses the role of animation in improving hit effects. In the game feel survey \cite{pichlmair}, the author emphasizes the function of hit stop in signifying the gameplay-relevant events and classifies hit stop as the research topic of impact feedback visualization. However, previous works mostly take an analysis of the games which are selected empirically. There is never a data-driven ranking for action games' performance of impact feel. Therefore, in this paper, we conduct a quantitative ranking and analysis by NLP. 

\textbf{Juicy Design.}
The word ``Juice" is often mentioned when discussing how to intensify interactivity. ``Juice" amplifies interactivity by providing excessive amounts of feedback in relation to user input \cite{jesper}. The goal of juice is to make actions feel significant and the results can be predicted unambiguously \cite{hicks2}. As Swink \cite{Swink} said, ``a juicy design can contribute to good game feel." However, too much juicy makes it hard to distinguish what aspects of interactivity have mechanical importance \cite{doucet}. Kao \cite{KAO} finds that none and extreme amounts of juiciness significantly decrease player experience. Because of the strong connection between juicy design and game feel, designers should study features of juicy impact feedback to learn how to enhance impact feel. Kieran et al. \cite{hicks1} explores game developers' understanding of game feel and juiciness through an online survey. They propose an empirically grounded framework for juicy design. The framework is applied to two commercial games and has proven helpful in analyzing game juiciness. In this paper, based on their framework, we further the study on juicy impact feedback design and figure out several features in action games that may influence players' feeling of impact.

\section{Methodology}
\subsection{Overview}
In this section, we conduct a quantitative analysis of players' comments on popular action games to find those that perform best in providing a satisfying impact feel. First, we select games from \textit{Steam} as our data set and collect players' comments. We manually annotate 5000 comments by their relevance to impact feel and train a NLP model. By the model, we obtain a list of games whose comments are highly dependent on the performance of impact feel and rank them. Afterward, we propose a 19-feature impact feedback design framework and apply it to our research target games.

\subsection{Data Collection and Ranking}
The first step is to collect comments that are related to impact feel. \textit{Steam} is one of the most popular digital game delivery platforms. This data-driven platform can indicate players' preferences for games and show comments. Previous work has proved that players' reviews can provide developers with useful information \cite{lindayi}. Therefore, we assume \textit{Steam}'s top seller list with tag ``fighting" can represent popular games that highly focus gameplay on fighting. 
To avoid influence from the preference system, we log out of the steam account and search under the visitor mode. Among the 9000 games on the top seller list, only 115 games have the tag ``fighting". It includes three genres: Fighting game, Beat 'em up, and Action role-playing game (ARPG). We remove games with less than 50 comments to ensure reliability and get 96 games. 

In the English player's community, there are several expressions for impact feel, such as ``the feeling of hit" \cite{moon} and ``combat impact" \cite{song}. It makes the training of NLP model difficult. However, there is an explicit Chinese word clearly describing impact feel. When Chinese players try to express how satisfying the impact feels, the terms tend to show convergence. Despite the difference in language and culture, players worldwide share a similar game feel. Therefore, we choose Chinese as our target language to simplify the model training process. Our open-source dataset \footnote{https://github.com/linzhonghao3/ImpactFeelCommentDB} contains 281720 Chinese comments on the 96 games.

Naive Bayesian is usually applied to text classification \cite{Zhang}. In this paper, we adapt this algorithm to classify comments into two categories, 0 represents meaningless comments and 1 illustrates  comments related to impact feel. Two researchers are responsible for manually classifying the comments. Both of them have game experience in action games for more than 500 hours, so they are qualified to distinguish whether a comment is discussing impact feel. We randomly choose 5000 comments from 15 of the 96 games and label them. 1000 comments are randomly chosen as the test set, and the others (4000) are used as the train set. We observe that only 400 of 4000 comments in the train set are classified as category 1, which leads to an asymmetry. To avoid the imbalance, after several tries, we finally apply the ratio 3:1 to our train set, which is composed of 1200 unrelated comments and 400 related comments. The trained model achieves an accuracy rate of 88$\%$ for category 0 and 92$\%$ for category 1 in the test set.

The model is then applied to the 280K comments of the chosen 96 games. We calculate the coefficient ``Impact Feel (IF) relevance" to represent the game's relevance with impact feel. A higher value means more players have ever commented on the game's impact feel. We collect the other coefficient ``Impact Feel (IF) attitude" by the positive rate of impact-feel related comments to represent players' attitude toward the impact feel. Over the 96 games, there are respectively 93, 85, 44 games with the ``IF relevance" larger than 5$\%$, 10$\%$, and 20$\%$. We set 20$\%$ as the threshold value for ``IF relevance" to further truncate games with a weak correlation of impact feel. Finally, we obtained 44 popular games whose comments are highly related to impact feel. We select the top and last 8 (20$\%$) games as our research targets to compare the difference.

\subsection{Features of Juicy Design in Combat System}
As Swink \cite{Swink} said, ``a juicy design can contribute to a good game feel." Therefore, the performance of impact feel largely depends on the impact feedback design. Our goal is to find detailed features of impact feedback design that leads to the difference in players' attitude toward impact feel. As shown in Table 1, by applying Kieran et al.'s juicy design framework \cite{hicks1} to chosen games, we furthered the framework to detailed technologies contributing to impact feel. For example, the concept of Focus of Attention in juicy design is detailed as On-Hit Effect and Camera Control. 

Based on our analysis of the 16 games, we identify 19 features currently used in impact feedback design to improve the impact feel. This section will present each feature's definitions, examples, and interesting findings.

\newcommand{\tabincell}[2]{\begin{tabular}{@{}#1@{}}#2\end{tabular}}
\begin{table}[b!]
\centering
\vspace{-1em}
  \footnotesize
  \begin{tabular}{ll}
    \toprule[2pt]
    Kieran's framework factor& Features on impact feedback\\
    \midrule[1.5pt]
    \textbf{A. Game Characteristics}  &  \\
    \hline
    A1. Mechanic 
    & \tabincell{c}{A1.1 Game Mechanic}\\
    \hline
    A2. Thematic Coherence
    & A2.1 Story Background\\
    \hline
    A3. Gameplay Coherence
    & A3.1 Combat system\\
    \hline
    A4. Feedback Coherence
    & A4.1 Impact Feedback System \\
    \midrule[1.5pt]
    \textbf{B. Game State} & \\
    \hline
    B1. Exaggerate 
    & \tabincell{l}{B1.1 Attack Effect\\B1.2 After-Hit Effect\\B1.3 Camera Effect
    \\B1.4 Hit Stop\\B1.5 Slow Motion}\\
    \hline
    B2. Focus of Attention 
    & \tabincell{l}{B2.1 On-Hit Effect\\B2.2 Camera Control}\\
    \hline
    B3. Highlighting: 
    & \tabincell{l}{B3.1 Color flashing}\\ 
    \hline
    B4. Ambient Feedback
    & \tabincell{l}{B4.1 Background Breathing}\\
    \midrule[1.5pt]
    \textbf{C. Direct Feedback}&\\
    \hline
    C1. Confirmatory 
    & \tabincell{l}{C1.1 Animation Switching}\\
    \hline
    C2. Multimodal
    & \tabincell{l}{C2.1 Weapon Sound Effect\\ C2.2 Character Sound Effect}\\
    \hline
    C3. Unambiguous: 
    & \tabincell{l}{C3.1 Sound Coherence}\\
    \hline
    C4A. Relevant
    & C4A.1 Slight/Heavy Attack\\
    \hline
    C4B. Supplemental feedback
    &\tabincell{l}{C4B.1 User Interface}\\
  \bottomrule[2pt]
\end{tabular}
\caption{Features of juicy design in combat system}
\label{tab_feature}
\end{table}

\subsubsection{A. Game Characteristic}
The game type of our research targets is composed of fighting games, beat 'em up and ARPG. Based on the similar characteristic of the combat system, in this paper we classify ARPG as fighting games as well.

\textbf{A1. Mechanic}
Do actions translate to expected feedback?

\textit{A1.1 Game Mechanic.} 
Because of the difference in the objective of gameplay between fighting games and beat 'em up, the impact feedback system should be designed differently. For beat 'em up, the goal is to beat an amount of non-player enemies \cite{tao}. The satisfaction of killing spring is a tasty experience for players. Therefore, beat 'em up pay more attention to after-hit effects, such as body destruction, which builds an invincible feeling for players.

\textbf{A2. Thematic Coherence}
Are the world and reactions to events believable in the context of the game? 

\textit{A2.1 Story Background.} The story background can largely influence the thematic coherence of the combat system, including weapons, skills, and special effects. 

\textbf{A3. Gameplay Coherence}
Are the mechanics compatible with each other?

\textit{A3.1 Combat System.} The combat system should be well-designed to fit the coherence with its gameplay mode. For example, players repeatedly kill enemies on the battlefield in beat 'em up. The developer should provide a variety of attacks and a juicy combo system to reduce aesthetic fatigue and keep stimulating players. As a counter-example, \textit{Sacred Citadel}\footnote{https://store.steampowered.com/app/207930/} receives several negative comments about how boring the combat system is, for the minimal kind of attacks and monsters with too many health points.

\textbf{A4. Feedback Coherence}
Does feedback reflect the importance of the event?

\textit{A4.1 Impact feedback system.} The impact feedback system is responsible for providing feedback on every impact during the fighting process. The basic idea is the simulation of physics plus exaggeration. Simulation of physics can evoke human visceral emotions from daily life experiences, and exaggeration amplifies players' feelings. For example, the idea of Newton's law is deeply seated in our minds. You can push away a cup by applying force. The phenomena are always simulated in games as knocking back the foes by special attacks. The distance can be adjusted for exaggeration purpose. 

\subsubsection{B. Game State}
The game state changes continuously during the fighting process., the offensive and defensive state is swapping in every second. 

\textbf{B1. Exaggerate}
Are reactions to action exaggerated to detail state change?

\textit{B1.1 Attack Effect.} Designers apply special effects in attack animations to make hits more gorgeous and powerful, usually by visualizing the trace of the weapon. In reality, we can observe the trail of fast-moving objects because of the persistence of human vision. As the simulation of this phenomenon, trace visualization is always used to exaggerate the speed of attack, which indirectly magnifies the power of the hit. 

\textit{B1.2 After-Hit Effect.} When characters are being hit, apart from the change of animation, some after-hit effects are applied to the character, such as bleeding and partial body destruction. This design vastly amplifies the impact of hits by describing how painful the foes are. In addition, some effects will remain on the screen for several seconds, prolonging the sense of achievement from the last kill. We find that all of the five beat 'em up apply after-hit effect. For example, in \textit{One Finger Death Punch}\footnote{https://store.steampowered.com/app/264200/}, blood sprays from the on-hit enemy, while in \textit{Fight's N Rage}\footnote{https://store.steampowered.com/app/674520/}, the death of a monster leads to a blast, and its bones scatter and fall to the ground.

\textit{B1.3 Camera Effect.} Camera shake and post-processing are representative examples of camera effects. Camera shake is commonly used to communicate a significant event like explosion, taking damage, or high-impact actions. Previous work has found that a carefully selected easing function in a semantically significant direction communicates more than randomly moving the camera \cite{pichlmair}. As an example, in \textit{Guilty Gear Xrd: Revelator}\footnote{https://arcsystemworks.com/game/guilty-gear-xrd-revelator/}, the camera slightly shakes in vertical when the character is knocked down to the ground from the air. When the player receives a transverse slash, the camera moves back and forth horizontally. Post-processing technology is always used to signify the change of game state. In \textit{The King of Fighters XIII}, the screen repeatedly flashes between white and red to mention the whole game is over. 

\begin{figure}
    \vspace{-1.0em}
    \centering
    \subfloat[Frame 1]{
        \includegraphics[width=0.47\linewidth]{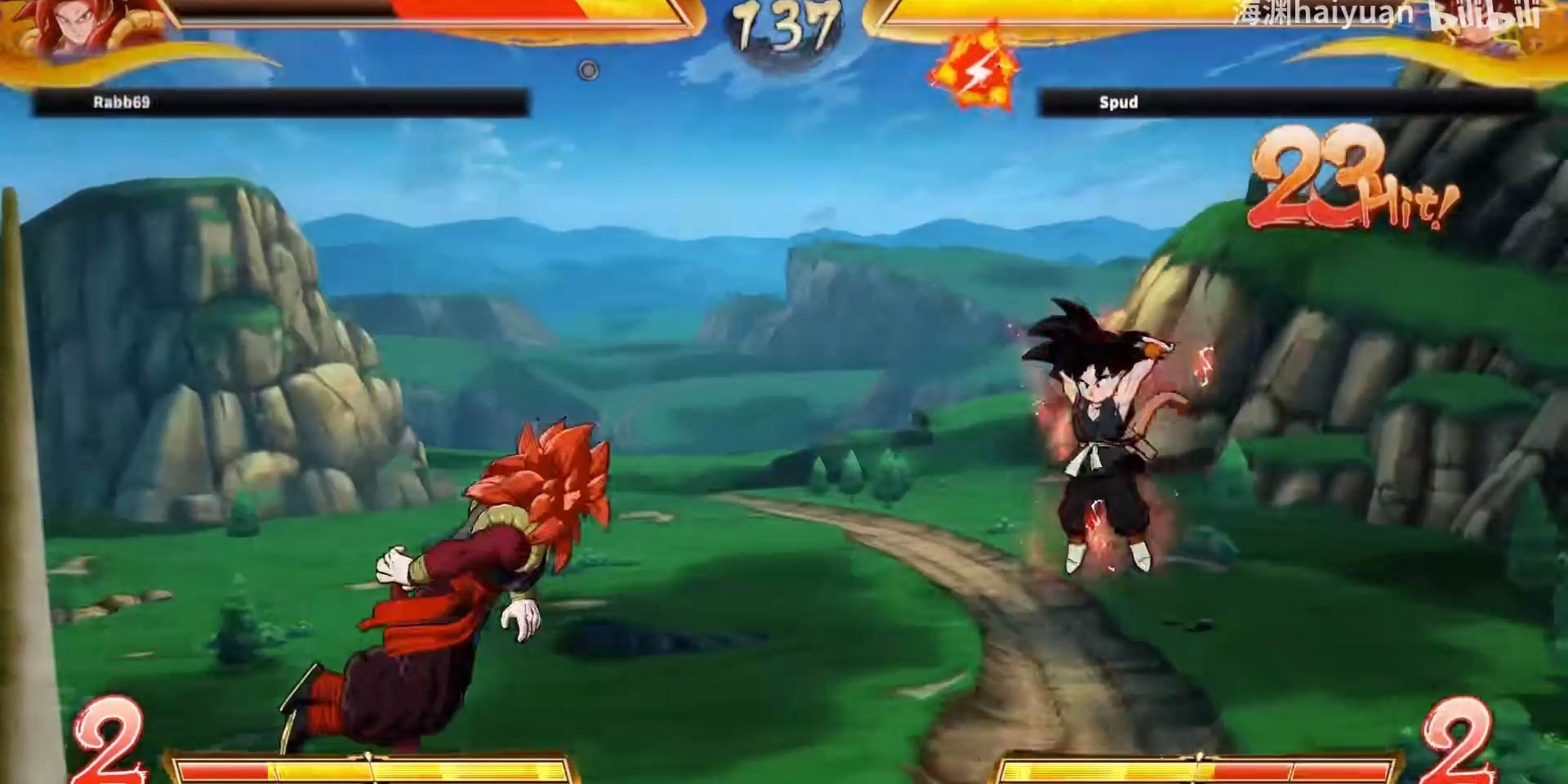} }
    \hfill
    \subfloat[Frame 31]{
        \includegraphics[width=0.47\linewidth]{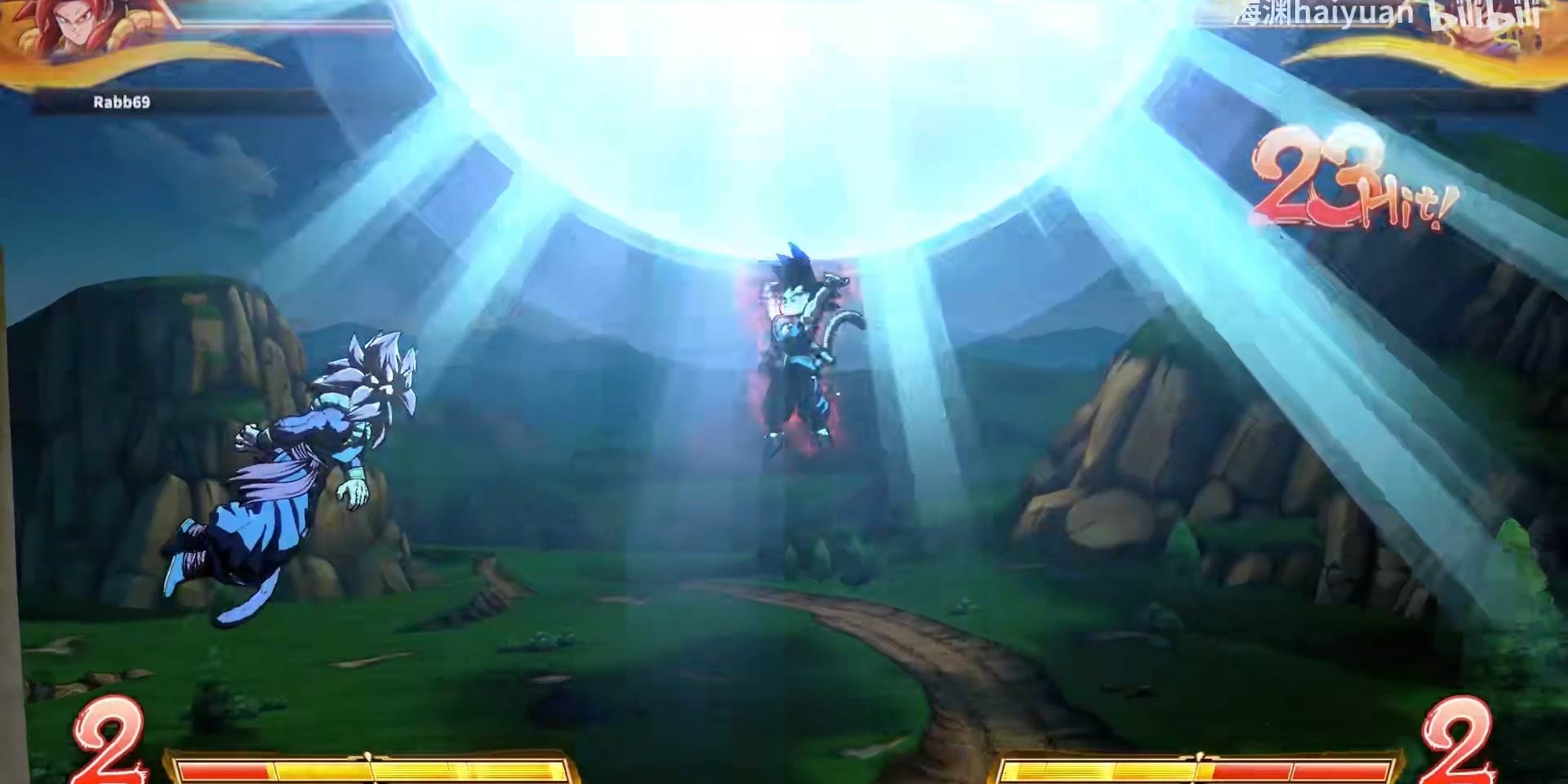} }
    \vspace{-0.5em}
    \caption{Hit stop in Dragon Ball FighterZ}
    \label{hit stop fig}
\end{figure}

\textit{B1.4 Hit Stop.} Basically, hit stop is to freeze the animation at the moment of impact. It is usually introduced to communicate feedback about the severity of a hit, but it can go further than that \cite{pichlmair}. Hit stop can be used in a way that distinguishes the weight and impact of a hit. For example, freeze small taps for fast and light combos to create rhythm and freeze more frames for a pumped attack to amplify the power \cite{hitstop}. In \textit{Guilty Gear Xrd -SIGN-}\footnote{https://arcsystemworks.com/game/guilty-gear-xrd/}, light attacks freeze the animation for 7 frames while the heavy attacks freeze for about 10 frames. In \textit{Dragon Ball FighterZ}\footnote{https://www.bandainamcoent.com/games/dragon-ball-fighterz}, the opponent's animation will freeze for a whole second once the GoKu (character name) is accumulating for his ultimate skill. This is shown in Figure \ref{hit stop fig}.

\begin{figure}
    \vspace{-1.5em}
    \centering
    \subfloat[Spot effect]{
        \includegraphics[width=0.47\linewidth]{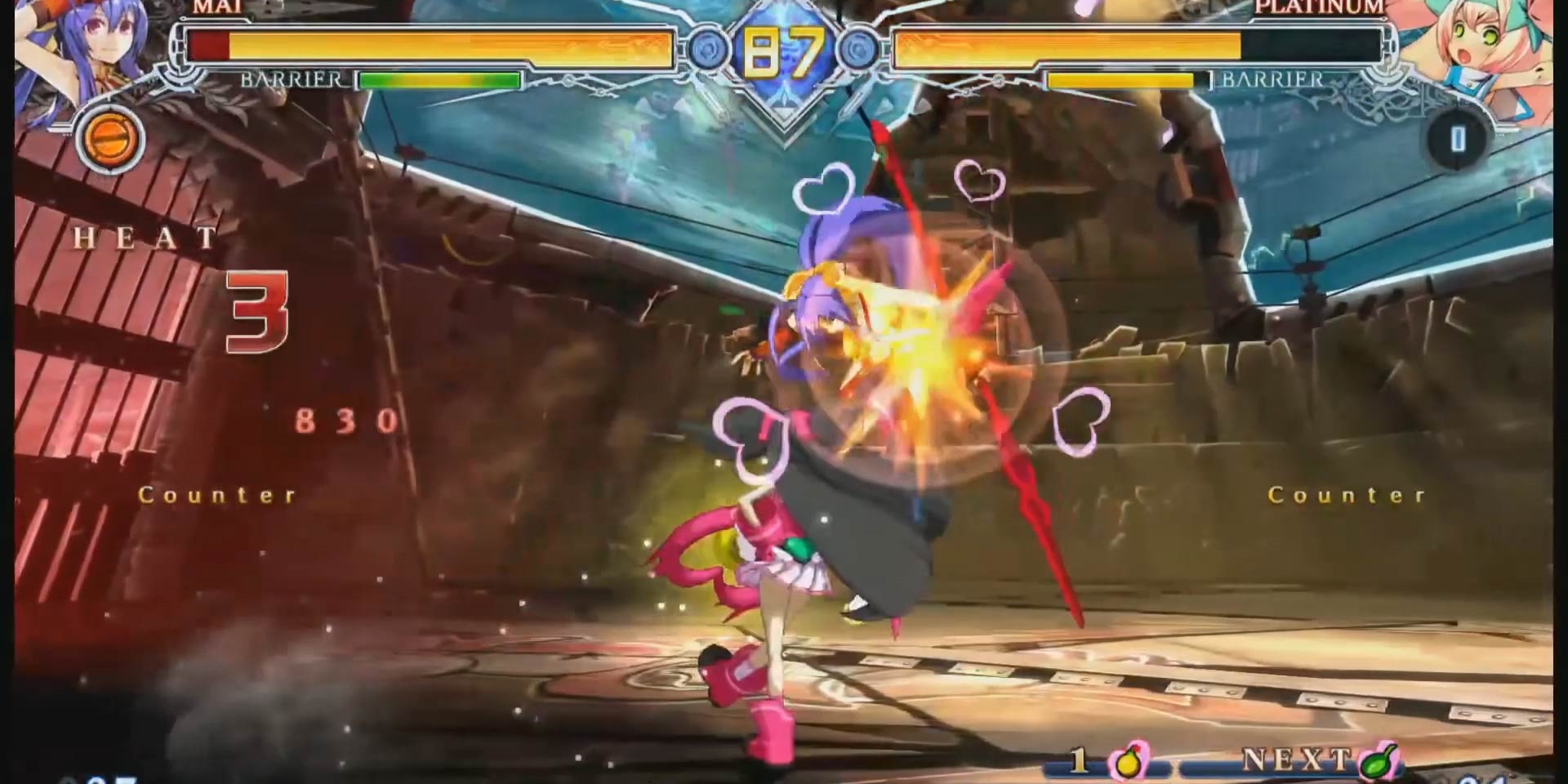}}
    \hfill
    \subfloat[Direction effect]{
        \includegraphics[width=0.47\linewidth]{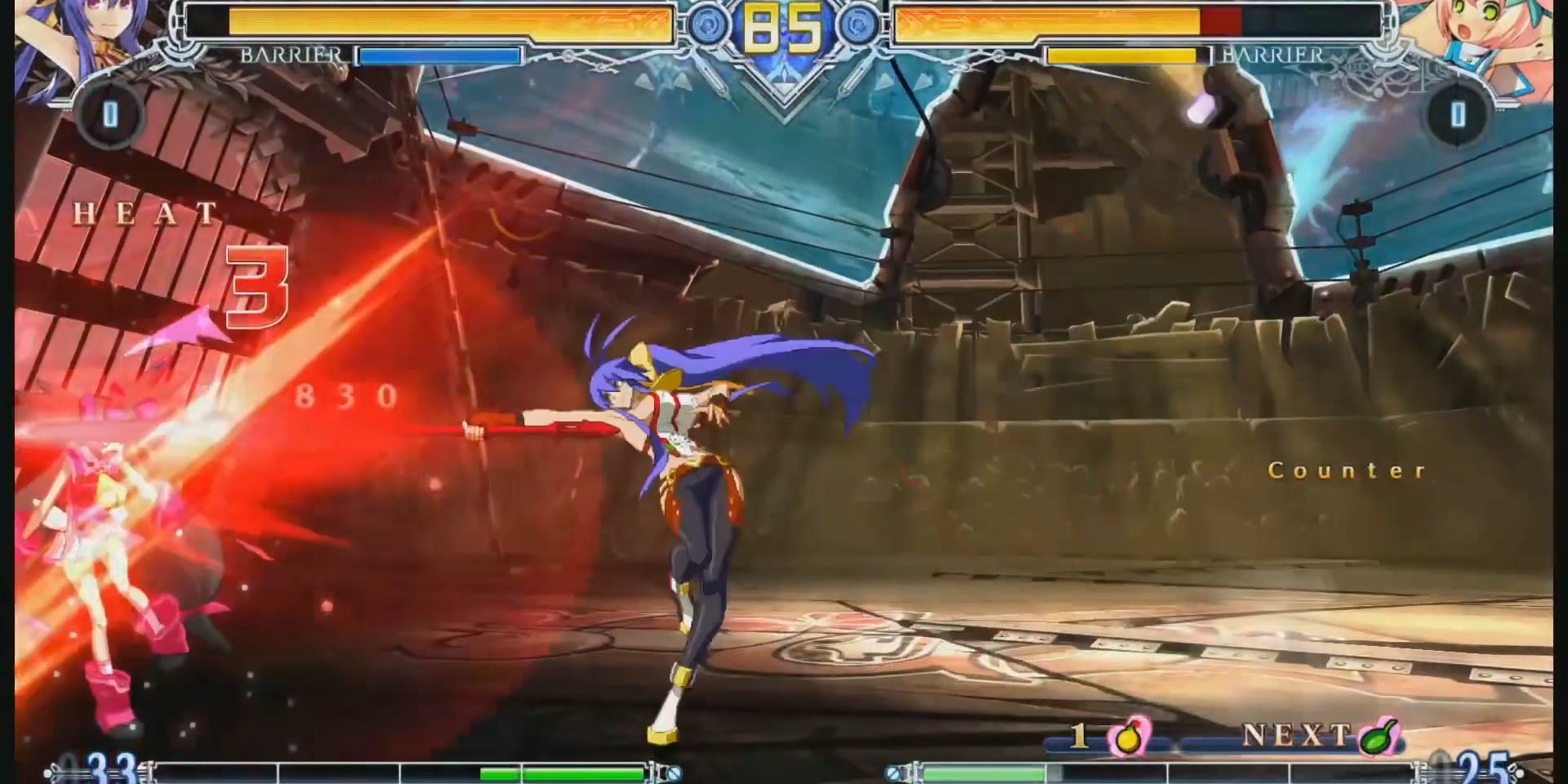}}
    \vspace{-0.5em}
    \caption{On-hit effect in BlazBlue Centralfiction}
    \vspace{-1.5em}
    \label{on-hit effect fig}
\end{figure}

\begin{figure*}[!h]
\centering
 \vspace{-1em}
    \label{animation switch fig}
    \subfloat[Frame 1]{\includegraphics[width=0.24\linewidth]{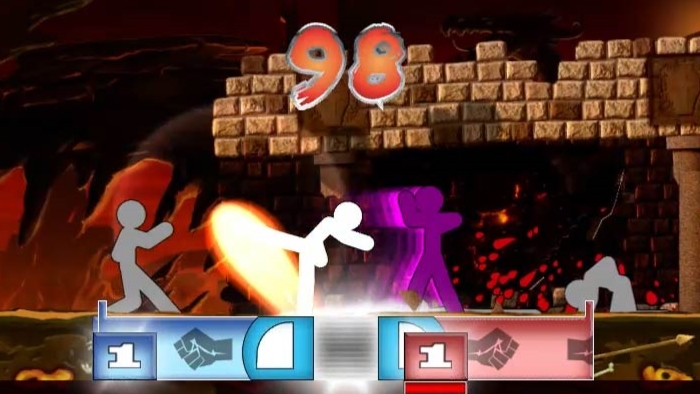}}
    \hfill
    \subfloat[Frame 2]{\includegraphics[width=0.24\linewidth]{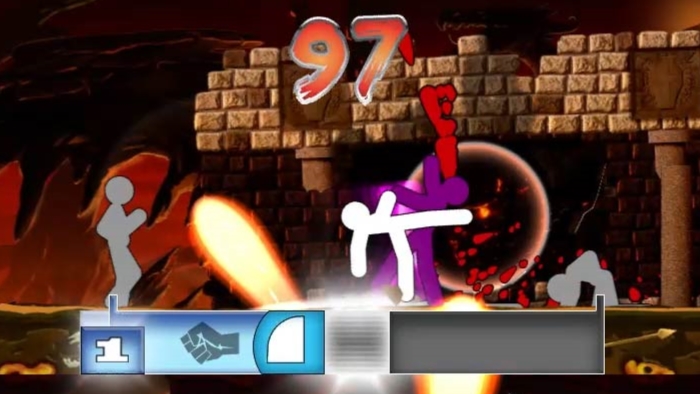}}
    \hfill
    \subfloat[Frame 3]{\includegraphics[width=0.24\linewidth]{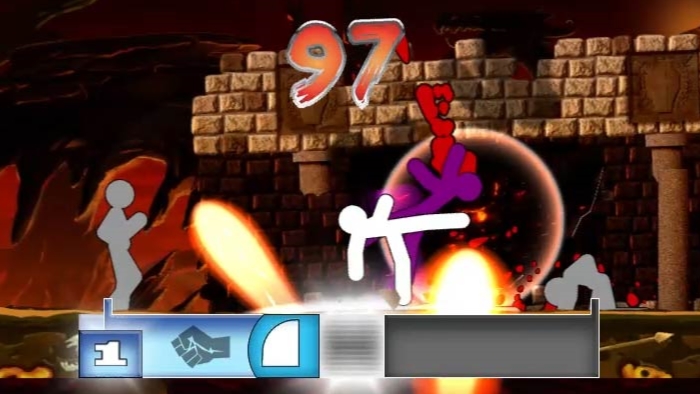}}
    \hfill
    \subfloat[Frame 5]{\includegraphics[width=0.24\linewidth]{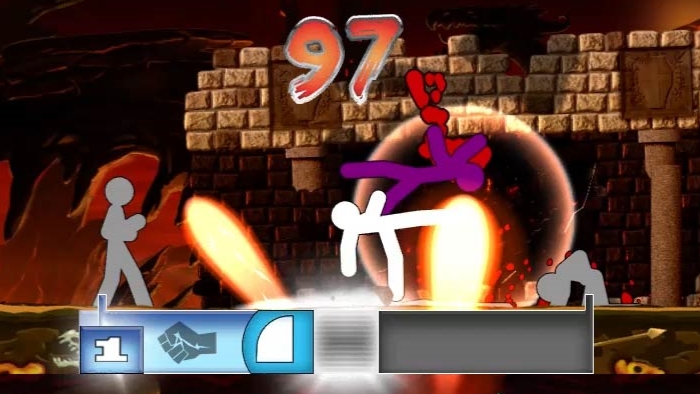}}
    \vspace{-0.5em}
    \caption{Animation Switch in One Finger Death Punch}
    \vspace{-1.5em}
\end{figure*}

\textit{B1.5 Slow Motion.} Slow motion is to slow down the time flow in a short duration. It is usually employed at the finishing attack. The slow motion makes the climactic finishing attacks clearer to the viewer \cite{song}. In \textit{One Finger Death Punch}, slow motion is applied to the fatality that fetches the foe's heart.

\textbf{B2. Focus of Attention.} 
Does the game feature feedback elements that draw your attention?

\textit{B2.1 On-Hit effect.} The difference to after-hit effect, described above, is that those are effects applied to communicate the impact of your strike. While the on-hit effects are spot-pattern effects that try to highlight the hit's position. Besides, on-hit effect can express more than just pointing out the impact position. Some games will apply on-hit effect with different patterns to distinguish the armaments used. Figure \ref{on-hit effect fig} shows that the on-hit effect varies for various weapons in \textit{BlazBlue Centralfiction}\footnote{https://arcsystemworks.com/game/blazblue-centralfiction-special-edition/}. The spot pattern is employed for fist and blunt objects, while an extra directional effect is applied when sharp weapons like spear and sword have slashed at the target.

\textit{B2.2 Camera Control.} The control of camera is an effective way to draw player's attention to significant events like ultimate skills and the finishing attack. Take the scene in Figure \ref{hit stop fig} as an example; when Goku is accumulating the ultimate skill “Spirit Bomb”, the camera gradually zooms in toward the growing blue ball. When the bomb is ready, the camera captures the smiling face of Goku, stays for several frames, and then quickly zooms out to show how this powerful attack hits the opponent. The dedicated camera control provides a fantastic user experience that is described as ``it is like watching a movie more than playing a game".

\textbf{B3. Highlighting} 
Are feedback elements that highlight information in harmony with other systems?

\textit{B3.1 Color flashing.} Color flashing is commonly used in combat systems to highlight the change of game state. The most pervasive method is to flash the character when on hit. 

\textbf{B4. Ambient Feedback} 
Is there feedback about the state of the world that is available without explicit player input, making the world appear natural and interactive?

\textit{B4.1 Background Breathing.} Background elements are breathing without an explicit input, such as the swaying grasses, the falling leaves, and the flowing water.

\subsubsection{C. Direct Feedback}
Direct feedback is significant in improving the impact feel. Elaborate feedback can respond immediately to the input, making an interactive experience. 

\textbf{C1. Confirmatory} 
Does the game give a direct response to physical input of a button?

\textit{C1.1 Animation Switching.} Animation switching is the technology used to blend between two animations. Fighting games attract players with quick tempo juicy battles, which provide a variety of attack patterns to choose from. Therefore, how to switch animation properly in a short period becomes a critical problem in game development. Most games apply a smooth lerping between predefined animations and idle animation. 
However, we are surprised to find the other idea of switching without any interpolation in \textit{One Finger Death Punch}, which won the highest favorite rate among the 44 games. The idea is to switch the pose directly, as you can see from frame 1 and 2 in Figure 3, the character instantly changes its posture without any transition. During the period from frame 2 to frame 5, the animation of the character itself does not change at all, while the movement of the attack effect indicates the direction of this kicking and makes the action understandable.

\textbf{C2. Multimodal} 
Is feedback for one action presented on multiple channels at once? (visual, audio)
The haptic part is not discussed in this paper since all the target games are running on PC platform which does not originally support haptic feedback without extra peripherals. 

\textit{C2.1 Weapon Sound Effect.} Sound effects are applied to weapons to provide audio feedback to attacks. It contributes to focusing the player's attention by communicating information. A juicy weapon sound effect system should distinguish sound from different weapons, from missing and hitting, and even from different on-hit materials. Furthermore, to avoid auditory fatigue, some games will prepare several sound clips for a frequently used attack and play one of them randomly.

\textit{C2.2 Character Sound Effect.} During the fighting process, the roar of the attacker and the scream of the foes can heat up the battle by amplifying audio feedback.

\textbf{C3. Unambiguous} 
Can information be connected to actions and only interpreted in one way?

\textit{C3.1 Sound Coherence.} 
Disharmony between vision and sound, such as a time delay between observing visual feedback and capturing the hit sound effect, will lead to a sense of unnaturalness and ruins the user experience.

\textbf{C4A. Relevant} Is feedback given in response to game critical events or is feedback received on minor player actions that require no further action.

\textit{C4A.1 Slight/Heavy Attack.} The feedback for light and heavy attacks should be distinguished. For light attacks, meaningful, direct feedback is suitable for communicating necessary information to players. For feedback to heavy attacks, more decorative elements can be applied to improve the visual effect and reward the player.

\textbf{C4B. Supplementary feedback} Does the game offer subtle additional feedback to emphasize actions already communicated in other ways, or minor player actions?

\textit{C4B.1 User interface.} The most correlated element in the user interface to combat system is the health bar and damage number. For the damage number, some games will turn the damage number bigger and color it red when a critical hit happens. For the health bar, the loss of health can be visualized by the gradual reduction or direct loss. 

\begin{table*}[!t]
    \setlength\tabcolsep{3pt}
    \scriptsize
    \resizebox{\linewidth}{!}{
    \begin{tabular}{c|c|c|ccccc|cc|c|c|c|cc|c|c|c p{10pt}}
    \hline
         \multirow{2}*{(A)Genre} & \multirow{2}*{Rank}  & \multirow{2}*{\diagbox{Game Names}{Game Features}}
          & \multicolumn{5}{c|}{B1} & \multicolumn{2}{c|}{B2}& \multicolumn{1}{c|}{B3} & \multicolumn{1}{c|}{B4} & C1 & \multicolumn{2}{c|}{C2}& {C3}&{C4A}&\multicolumn{1}{c}{C4B}\\
    \cline{4-19}
         & & & B1.1 & B1.2 & B1.3 & B1.4 & B1.5 & B2.1 & B2.2 & B3.1 & B4.1 & C1.1 & C2.1 & C2.2 & C3.1 & C4A.1 & C4B.1 \\
    \hline
        \multirow{11}* {\rotatebox{90}{Fighting Games}} &
        2& GUILTY GEAR Xrd: REVELATOR & \textcolor{green}{\checkmark} & \textcolor{red}{\ding{53}} & \textcolor{green}{\checkmark} & \textcolor{green}{\checkmark} & \textcolor{green}{\checkmark} & \textcolor{green}{\checkmark} & \textcolor{green}{\checkmark} & \textcolor{green}{\checkmark} & \textcolor{green}{\checkmark} & \textcolor{green}{\checkmark} & \textcolor{green}{\checkmark} & \textcolor{green}{\checkmark} & \textcolor{green}{\checkmark} & \textcolor{green}{\checkmark} & \textcolor{green}{\checkmark} \\ 
        
        & 3& BlazBlue: Centralfiction & \textcolor{green}{\checkmark} & \textcolor{red}{\ding{53}} & \textcolor{green}{\checkmark} & \textcolor{green}{\checkmark} & \textcolor{green}{\checkmark} & \textcolor{green}{\checkmark} & \textcolor{green}{\checkmark} & \textcolor{green}{\checkmark} & \textcolor{green}{\checkmark} & \textcolor{green}{\checkmark} & \textcolor{green}{\checkmark} & \textcolor{green}{\checkmark} & \textcolor{green}{\checkmark} & \textcolor{green}{\checkmark} & \textcolor{green}{\checkmark} \\
        
        & 4& GUILTY GEAR Xrd -SIGN- & \textcolor{green}{\checkmark} & \textcolor{red}{\ding{53}} & \textcolor{green}{\checkmark} & \textcolor{green}{\checkmark} & \textcolor{green}{\checkmark} & \textcolor{green}{\checkmark} & \textcolor{green}{\checkmark} & \textcolor{green}{\checkmark} & \textcolor{green}{\checkmark} & \textcolor{green}{\checkmark} & \textcolor{green}{\checkmark} & \textcolor{green}{\checkmark} & \textcolor{green}{\checkmark} & \textcolor{green}{\checkmark} & \textcolor{green}{\checkmark} \\
        
        & 5& THE KING OF FIGHTERS XIII & \textcolor{green}{\checkmark} & \textcolor{red}{\ding{53}} & \textcolor{green}{\checkmark} & \textcolor{green}{\checkmark} & \textcolor{green}{\checkmark} & \textcolor{green}{\checkmark} & \textcolor{green}{\checkmark} & \textcolor{green}{\checkmark} & \textcolor{green}{\checkmark} & \textcolor{green}{\checkmark} & \textcolor{green}{\checkmark} & \textcolor{green}{\checkmark} & \textcolor{green}{\checkmark} & \textcolor{green}{\checkmark} & \textcolor{green}{\checkmark} \\
        
        & 8& DRAGON BALL FighterZ & \textcolor{green}{\checkmark} & \textcolor{red}{\ding{53}} & \textcolor{green}{\checkmark} & \textcolor{green}{\checkmark} & \textcolor{green}{\checkmark} & \textcolor{green}{\checkmark} & \textcolor{green}{\checkmark} & \textcolor{green}{\checkmark} & \textcolor{green}{\checkmark} & \textcolor{green}{\checkmark} & \textcolor{green}{\checkmark} & \textcolor{green}{\checkmark} & \textcolor{green}{\checkmark} & \textcolor{green}{\checkmark} & \textcolor{green}{\checkmark} \\
        
        \cline{2-19}
        
        & 37&DEAD OR ALIVE 5 & \textcolor{red}{\ding{53}} & \textcolor{red}{\ding{53}} & \textcolor{green}{\checkmark} & \textcolor{green}{\checkmark} & \textcolor{green}{\checkmark} & \textcolor{green}{\checkmark} & \textcolor{green}{\checkmark} & \textcolor{green}{\checkmark} & \textcolor{green}{\checkmark} & \textcolor{green}{\checkmark} & \textcolor{green}{\checkmark} & \textcolor{green}{\checkmark} & \textcolor{green}{\checkmark} & \textcolor{green}{\checkmark} & \textcolor{green}{\checkmark} \\
        
        & 38&BlazBlue: Cross Tag Battle & \textcolor{green}{\checkmark} & \textcolor{red}{\ding{53}} & \textcolor{green}{\checkmark} & \textcolor{green}{\checkmark} & \textcolor{green}{\checkmark} & \textcolor{green}{\checkmark} & \textcolor{green}{\checkmark} & \textcolor{green}{\checkmark} & \textcolor{green}{\checkmark} & \textcolor{green}{\checkmark} & \textcolor{green}{\checkmark} & \textcolor{green}{\checkmark} & \textcolor{green}{\checkmark} & \textcolor{green}{\checkmark} & \textcolor{green}{\checkmark} \\
        
        & 41&Hunter's arena: legends & \textcolor{green}{\checkmark} & \textcolor{red}{\ding{53}} & \textcolor{green}{\checkmark} & \textcolor{green}{\checkmark} & \textcolor{green}{\checkmark} & \textcolor{green}{\checkmark} & \textcolor{green}{\checkmark} & \textcolor{green}{\checkmark} & \textcolor{green}{\checkmark} & \textcolor{green}{\checkmark} & \textcolor{green}{\checkmark} & \textcolor{green}{\checkmark} & \textcolor{green}{\checkmark} & \textcolor{green}{\checkmark} & \textcolor{green}{\checkmark} \\
        
        & 42&DISSIDIA FINAL FANTASY NT & \textcolor{green}{\checkmark} & \textcolor{red}{\ding{53}} & \textcolor{green}{\checkmark} & \textcolor{green}{\checkmark} & \textcolor{green}{\checkmark} & \textcolor{green}{\checkmark} & \textcolor{green}{\checkmark} & \textcolor{green}{\checkmark} & \textcolor{green}{\checkmark} & \textcolor{green}{\checkmark} & \textcolor{green}{\checkmark} & \textcolor{green}{\checkmark} & \textcolor{green}{\checkmark} & \textcolor{green}{\checkmark} & \textcolor{green}{\checkmark} \\
        
        & 43&NARUTO: Ultimate Ninja STORM & \textcolor{green}{\checkmark} & \textcolor{red}{\ding{53}} & \textcolor{green}{\checkmark} & \textcolor{green}{\checkmark} & \textcolor{green}{\checkmark} & \textcolor{green}{\checkmark} & \textcolor{green}{\checkmark} & \textcolor{green}{\checkmark} & \textcolor{green}{\checkmark} & \textcolor{green}{\checkmark} & \textcolor{green}{\checkmark} & \textcolor{green}{\checkmark} & \textcolor{red}{\ding{53}} & 
        \textcolor{green}{\checkmark} & \textcolor{green}{\checkmark}  \\
        
        & 44&ONE PUNCH MAN & \textcolor{green}{\checkmark} & \textcolor{red}{\ding{53}} & \textcolor{green}{\checkmark} & \textcolor{green}{\checkmark} & \textcolor{green}{\checkmark} & \textcolor{green}{\checkmark} & \textcolor{green}{\checkmark} & \textcolor{green}{\checkmark} & \textcolor{green}{\checkmark} & \textcolor{green}{\checkmark} & \textcolor{green}{\checkmark} & \textcolor{green}{\checkmark} & \textcolor{red}{\ding{53}} & 
        \textcolor{green}{\checkmark} & \textcolor{green}{\checkmark} \\
        \hline
        \multirow{5}*{\rotatebox{90}{Beat 'em up}} &
        1&ONE Finger Death Punch & \textcolor{green}{\checkmark} & \textcolor{green}{\checkmark} & \textcolor{green}{\checkmark} & \textcolor{green}{\checkmark} & \textcolor{green}{\checkmark} & \textcolor{green}{\checkmark} & \textcolor{green}{\checkmark} & \textcolor{red}{\ding{53}} & \textcolor{green}{\checkmark} & \textcolor{green}{\checkmark} & \textcolor{green}{\checkmark} & \textcolor{green}{\checkmark} & \textcolor{green}{\checkmark} & \textcolor{green}{\checkmark} & \textcolor{green}{\checkmark}  \\
        
        & 6&9 Monkeys of Shaolin & \textcolor{green}{\checkmark} & \textcolor{green}{\checkmark} & \textcolor{green}{\checkmark} & \textcolor{green}{\checkmark} & \textcolor{green}{\checkmark} & \textcolor{green}{\checkmark} & \textcolor{green}{\checkmark} & \textcolor{green}{\checkmark} & \textcolor{green}{\checkmark} & \textcolor{green}{\checkmark} & \textcolor{green}{\checkmark} & \textcolor{green}{\checkmark} & \textcolor{green}{\checkmark} & \textcolor{green}{\checkmark} & \textcolor{green}{\checkmark}  \\
        
       & 7&Fight's N Rage & \textcolor{green}{\checkmark} & \textcolor{green}{\checkmark} & \textcolor{green}{\checkmark} & \textcolor{green}{\checkmark} & \textcolor{green}{\checkmark} & \textcolor{green}{\checkmark} & \textcolor{green}{\checkmark} & \textcolor{green}{\checkmark} & \textcolor{green}{\checkmark} & \textcolor{green}{\checkmark} & \textcolor{green}{\checkmark} & \textcolor{green}{\checkmark} & \textcolor{green}{\checkmark} & \textcolor{green}{\checkmark} & \textcolor{green}{\checkmark}  \\
        \cline{2-19}
        
        & 39&Sacred Citadel & \textcolor{green}{\checkmark} & \textcolor{green}{\checkmark} & \textcolor{green}{\checkmark} & \textcolor{red}{\ding{53}} & \textcolor{green}{\checkmark} & \textcolor{green}{\checkmark} & \textcolor{red}{\ding{53}} & \textcolor{green}{\checkmark} & \textcolor{green}{\checkmark} & \textcolor{green}{\checkmark} & \textcolor{green}{\checkmark} & \textcolor{green}{\checkmark} & \textcolor{green}{\checkmark} & \textcolor{green}{\checkmark} & \textcolor{green}{\checkmark}  \\
        
        & 40&Draw Slasher & \textcolor{green}{\checkmark} & \textcolor{green}{\checkmark} & \textcolor{red}{\ding{53}} & \textcolor{red}{\ding{53}} & \textcolor{red}{\ding{53}} & \textcolor{red}{\ding{53}} & \textcolor{red}{\ding{53}} & \textcolor{green}{\checkmark} & \textcolor{green}{\checkmark} & \textcolor{green}{\checkmark} & \textcolor{green}{\checkmark} & \textcolor{green}{\checkmark} & \textcolor{red}{\ding{53}} & 
        \textcolor{green}{\checkmark} & \textcolor{red}{\ding{53}} &\\
    \hline
    \end{tabular}}
     \label{tab_feature1}
    \caption{Distribution of the 15 features on the top 8 and last 8 games}
    \vspace{-2em}
\end{table*}

\section{Result}
By comparing the distribution of features on the top games and the last games, we figure out three elements that largely influence players' impact feel. They can be used as one of the evaluation methods to examine a game's impact feel. The results is provided in Table 2. They are respectively:

\textbf{1) Hit stop.} Almost all games apply hit stop in their combat system to create rhythm and distinguish attacks except \textit{Sacred Citadel} and \textit{Draw Slasher}, which rank 39th and 40th on the list.
The impact feedback is weak for the lack of hit stop, making the attacks soft and powerless. The boundary between missing and hitting, quick attack and heavy attack is indistinct. Without hit stop, the impact feedback system cannot communicate enough information for players to feel the impact. Players describe them as dull, with less than 70$\%$ ``IF attitude".

\textbf{2) Sound coherence.} The coherence of audio feedback is a critical element in impact feedback system. It contributes to the game atmosphere, which constitutes an important, but also highly nuanced, potentially subjective, and subtle experience
\cite{ribeiro}.
The broken coherence, such as improper sound effects and time delay between visual and audio feedback, can lead to a sense of unnaturalness, which essentially ruins the immersive experience. It is a reason for the low favorite rate of \textit{One Punch Man: A Hero Nobody Knows}. Several players complain about the monotonous background music and minimal kinds of sound effects. Therefore, maintaining the sound coherence is necessary for a juicy impact feedback design.

\textbf{3) Camera control.} Camera control is usually used to draw attention to significant events such as heavy attacks and finishing attacks. The distribution of Camera control varies in the five beat 'em up. All of the three with high rank apply, while none of the two with low rank apply. In these two games, the camera just simply moves following the character. The loss of dedicated camera control dulls the combat experience and weakens the feeling of impact.

\textbf{Extra Discovery.} During the analysis, we discovered an interesting phenomenon that fighting games have essentially all the features despite their ranking while beat 'em up do not. For beat 'em up, it is evident that games with high ranking applies more features than those with low ranking. A possible reason is that popular fighting games are mostly developed by well-known studios such as \textit{Arc System Works} and \textit{Team Ninjia}. Five of the eleven fighting games are developed by \textit{Arc System Works}, so the technologies applied are similar despite of the ranking. For beat 'em up, even small companies may create popular games. For example, \textit{One Finger Death Punch}, the first product of \textit{Sliver Dollar Games} wins the first prize in players' attitude toward its performance of impact feel. It reveals that the bar of developing a popular fighting game is higher than beat 'em up. It's hard for companies without prerequisite experience to produce popular fighting games. 

\section{Conclusion and future work}
Impact feel, a sub-genre of game feel in describing the sense of satisfaction from receiving the impact of hits, is a critical element in action games. However, very few researches have been done to provide a quantitative analysis from an academic perspective. In this paper, we conducted a data-driven analysis on ranking popular action games with their performance of impact feel by NLP. Based on previous work on juicy design, we deduced a 19-feature framework of impact feedback design. For each feature, we present the definition and exciting applications for better understanding. Afterward, we examine the framework on the top and last 8 games of our ranking. Therefore, we can explore the different distribution of features on games with excellent/awful impact feel. As a result, we identify that hit stop, sound coherence, and camera control play a significant role in influencing impact feel. The lack of even one of the three features will weaken the performance of impact feel and make it undesirable. We hope the result can be an evaluation metric for further study. 

There are a few limitations that need to be considered in our work. First, we only fed the data set of Chinese comments to the NLP model. Corpus with more languages should be considered to improve the generality of the text classifier. In addition, the small sample of games limited our result in finding more about the relationship between other features and the impact feel. We hope future work can extend our work into a larger sample that includes more action games. 

\section{Acknowledgment}
This work was supported by CUHK(SZ)-White Matrix Joint Metaverse Laboratory.

\bibliographystyle{IEEEtran}
\bibliography{reference}
\clearpage
\end{document}